\documentclass[11pt]{article}
\usepackage{geometry,epsf}
\usepackage{hyperref}
\def\gtwid{\mathrel{\raise.3ex\hbox{$>$\kern-.75em\lower1ex\hbox{$\sim
$}}}}
\def\vio{\mathrel{\hbox{$R$\kern-.60em\hbox{$/
$}}}}
\usepackage{graphicx}
\usepackage{amssymb}
\def\lsim{\mathrel{\raise.3ex\hbox{$<$\kern-.75em\lower1ex\hbox{$\sim$}}}}
\def\gsim{\mathrel{\raise.3ex\hbox{$>$\kern-.75em\lower1ex\hbox{$\sim$}}}}

\newcommand{\fr}[2]{{\hbox{$ #1 \over #2 $}}}

\begin{document}

\title
{Bremsstrahlung in dark matter annihilation}
\author{Vernon Barger$^{1}$, Wai-Yee Keung$^{2}$, Danny Marfatia$^{3,1}$\\[2ex]
\small\it ${}^{1}$Department of Physics, University of Wisconsin, Madison, WI 53706, U.S.A.\\
\small\it ${}^{2}$Department of Physics, University of Illinois, Chicago, IL 60607, U.S.A.\\
\small\it ${}^{3}$Department of Physics and Astronomy, University of
Kansas, Lawrence, KS 66045, U.S.A.}

\date{}

\maketitle

\begin{abstract}
We show that the energy spectra from dark matter (DM) annihilation into $f^+f^-\gamma$ (where $f$ is a light fermion) 
via chirality preserving interactions are identical in 3 scenarios: (a) DM is a Majorana fermion and the particle exchanged is a scalar, (b) DM is a Majorana fermion and the particle exchanged is a vector and (c) DM is a scalar and the particle exchanged is a fermion. For cases (a) and (c), we also
calculate the differential cross section to $\ell^+f^-V$, where $V=W,Z,\gamma$, and $\ell$ is a light fermion that may or not be the same as $f$. The form of the cross section depends on whether the DM is a Majorana fermion or a scalar but in both cases its form is independent of $V$. 
\end{abstract}
\newpage


Intensive searches for dark matter (DM) via their annihilation signatures~\cite{Bertone} have spurred interest 
in the effects of electromagnetic and electroweak bremsstrahlung on the rates for particular final states and on 
the resulting spectra~\cite{brem,brem2,bremm}.

In this Letter, we show that for 3 different interactions that are chirality preserving and appear in popular scenarios, dark matter annihilation to $f^+f^-\gamma$  in the static limit
gives identical energy spectra. We then connect 
the common structure for the amplitude to an operator basis. Finally, for 2 of the 3 cases, 
we derive the differential cross section to $\ell^+f^-V$, where $V=W,Z, \gamma$ and find its form to be different 
for Majorana DM and scalar DM, but independent of $V$.

\section*{Electromagnetic bremsstrahlung} 

We consider the annihilation final state $e^+e^-\gamma$ for the interaction terms in the table below. 
The exchanged particles $S$, $W'$ and $E$ (collectively denoted by $X$) are taken to be negatively charged and
 heavier than the corresponding DM particle  $D=\chi, N$ and $\phi$. We assume that $D$ is a standard model singlet and
  its coupling to the electron to be chiral with the interaction involving $e_R$. For $e_L$, straightforward replacements can be made.

$$ \begin{array}{c|c|c|c}
\hline
\hbox{} & \hbox{Model 1} & \hbox{Model 2}  & \hbox{Model 3} \cr \hline
 & & & \\
\hbox{DM} &
\hbox{Majorana $\chi$} &
\hbox{Majorana $N$  } & 
\hbox{Scalar  $\phi$ } \cr  & & & \\
 & & & \\
\hbox{Exchanged particle X}  &
\hbox{Scalar $S$}         &
\hbox{Vector $W'$}        &
\hbox{Fermion $E$}     \cr  & & & \\ 
 & & & \\
\hbox{Interaction }\ {\cal L}_I  &
g'S^\dagger \bar \chi e_R &
g'W'^\dagger_\mu \bar N \gamma^\mu e_R &
g' \bar E\phi e_R \cr  & & & \\  \hline 
\end{array}  
\label{table} $$
\\

The  supersymmetric case of neutralino DM with selectron exchange is a good example of  \mbox{Model 1}.
The case of a new heavy charged $W'$ gauge boson that couples $e_R$ and
a right-handed heavy Majorana neutrino $N$ in variations of left-right theory  
falls into the class of \mbox{Model 2}~\cite{Ma}.
Scenarios of scalar DM 
which have been of interest recently are realizations of \mbox{Model 3}~\cite{bremm}.

For the annihilation process, 
$DD \to e(p_1)  \bar e(p_2)  \gamma(k, \epsilon)$, in the nonrelativistic limit
we find  a unified description of the energy spectra of 
the decay products for all three models.
In standard notation, the amplitude turns out to be
\begin{equation}
{\cal M}=-  { 8eC\bar u_R(p_1)( \not p_2\not \epsilon \not k
                                  \  \mp \ \not k\not \epsilon \not p_1) v_R(p_2)
\over [(p_1-p_2+k)^2- 4m_X^2]
      [(p_1-p_2-k)^2- 4m_X^2] }  \,, 
      \label{amp}
\end{equation}
where the minus sign applies to Models 1 and 2 (with Majorana DM) and the plus sign applies to
Model 3 (with scalar DM).   
Note that for the Majorana DM
models only the initial state with vanishing total angular momentum participates in the
annihilation. The overall constant
$$ C=\left\{ 
\begin{array}{cc}
{i\over\sqrt{2}} g'^2 & \hbox{for Model 1, } \\
{i\over\sqrt{2}} g'^2 (2+ {m_D^2\over m_X^2} )  & \hbox{for Model 2, } \\
g'^2  &  \hbox{for Model 3. }\end{array}\right. $$
Remarkably, we find the same structure for all  3 Models even though Model 2 involves the tri-gauge boson vertex $W'W'\gamma$.
The amplitude obeys QED gauge invariance and its two terms do not
interfere when photon polarizations are summed in the limit $m_e \to 0$. 
The spin-averaged annihilation rate is
\begin{equation}
 v_{\rm rel}{d \sigma\over dx_1 dx_3}
= {1\over (2S_D+1)^2} \ 
{|eC|^2  \over 4\pi^3m_D^2}\
  \    {[(1-x_1)^2+(1-x_2)^2] (1-x_3)
                              \over
                               (1-2x_1-r)^2  (1-2x_2-r)^2  } \ ,
 \end{equation}                              
where $r=m_X^2/m_D^2$.  The scaling variables
$x_i=E_i/m_D$ ($i=1,2$), $x_3= E_\gamma/m_D$ are defined in the static center of mass frame so that
$x_1+x_2+x_3=2$. The spin-averaged factor $1\over (2S_D+1)^2$
is $\fr14$ for Models 1, 2 and unity for Model 3.
The photon energy distribution is obtained
by integrating over $x_1\in [1-x_3,1]$~\cite{brem}:
\begin{eqnarray}
 v_{\rm rel}{d \sigma\over dx_3}
&=& {1\over (2S_D+1)^2} \ 
{  {|eC|^2}  \over 32\pi^3m_D^2} \
 {1-x_3\over(1+r-x_3)^2} \nonumber \\
 &&\times
\left(   2 x_3{ x_3^2 + (1+r-x_3)^2 \over (1+r)(1+r-2x_3)}
-{(1+r)(1+r-2x_3)\over 1+r-x_3}\ln {1+r\over 1+r-2x_3} \right) \ .
\end{eqnarray}

The unified formulas for the amplitude and the distribution for the 3 models
are simple and interesting. 
We now study the structure of the amplitude. The
numerator of the amplitude involves two pieces, each of which are QED gauge invariant.
The first one is  
$$ {1\over 2}\bar u_R(p_1) \not p_2 \not {\epsilon} \not k v_R(p_2)
=\bar u_R(p_1)(p_2\cdot {\epsilon} \not k 
   - p_2 \cdot k \not {\epsilon} ) v_R(p_2) \longleftarrow {\cal O}
\ , $$ 
where we have used the massless fermion on-shell condition and
the last step identifies the stucture  with an operator~\cite{bremm},
$$  {\cal O} = F^{\mu\nu}   \bar\psi_R \gamma_\nu \partial_\mu\psi_R \ .   $$
Similarly, the second piece is 
$$ {1\over 2} \bar u_R(p_1) \not k \not {\epsilon} \not p_1 v_R(p_2)
=\bar u_R(p_1)(p_1\cdot {\epsilon} \not k - p_1 \cdot k \not {\epsilon} ) v_R(p_2) 
\longleftarrow {\cal O}^\dagger\ ,$$ 
where~\cite{bremm}
$$  {\cal O}^\dagger
= F^{\mu\nu}   (\partial_\mu \bar\psi_R) \gamma_\nu \psi_R  \ ,  $$
is the conjugate of ${\cal O}$.
On the other hand, we can use the Chisholm identity 
$$ \gamma^\alpha\gamma^\beta\gamma^\mu
=g^{\alpha\beta} \gamma^\mu -g^{\alpha\mu} \gamma^\beta
 +g^{\beta\mu} \gamma^\alpha 
-i{\epsilon}^{\alpha\beta\mu\nu}\gamma_\nu\gamma_5 \ ,$$
to write
\begin{eqnarray}
 \bar u_R(p_1) \not p_2 \not {\epsilon} \not k v_R(p_2)
&=&\bar u_R(p_1)( p_2\cdot {\epsilon} \not k -  p_2 \cdot k \not {\epsilon} 
           + \underbrace{k\cdot{\epsilon} \not p_2}_{\hookrightarrow 0} 
 -i{\epsilon}^{p_2{\epsilon}k\mu}\gamma_\mu\gamma_5 ) v_R(p_2)\nonumber \\ 
 &=&-2i {\epsilon}^{p_2{\epsilon}k\mu}\bar 
u_R(p_1) \gamma_\mu v_R(p_2)\,. \nonumber
\end{eqnarray}
Similarly,
$$ \bar u_R(p_1)  \not k \not {\epsilon}  \not p_1 v_R(p_2)   
=+2i {\epsilon}^{p_1{\epsilon}k\mu}
\bar u_R(p_1) \gamma_\mu  v_R(p_2)  \ .$$
So we have the relations,
$$  {\cal O}
= F^{\mu\nu}     \bar\psi_R \gamma_\nu \partial_\mu\psi_R    
= -i \widetilde{F}^{\mu\nu}    \bar\psi_R \gamma_\nu
\partial_\mu\psi_R   \ , $$
$$  {\cal O}^\dagger
= F^{\mu\nu}   (\partial_\mu \bar\psi_R) \gamma_\nu \psi_R 
= i\widetilde{F}^{\mu\nu}   (\partial_\mu \bar\psi_R) 
\gamma_\nu \psi_R  \,,   $$
where 
$\widetilde F^{\mu\nu}={1\over2}{\epsilon}^{\mu\nu\alpha\beta}F_{\alpha\beta}$.
Then,
$$   {\cal O}+{\cal O}^\dagger  =  F^{\mu\nu}  
\partial_\mu(\bar\psi_R \gamma_\nu \psi_R) 
\equiv {\cal F}   \ ,$$
$$ i({\cal O}-{\cal O}^\dagger)  =  \widetilde{F}^{\mu\nu}  
\partial_\mu(\bar\psi_R \gamma_\nu \psi_R)  
\equiv \widetilde {\cal F} \ .  $$
Thus, either ${\cal O}$ and ${\cal O}^\dagger$, or ${\cal F}$ and 
$\widetilde{\cal F}$ can be used as bases to describe the amplitude.
Because the dual substitution interchanges ${\bf E} \leftrightarrow {\bf  B}$, 
it also interchanges a right-polarized photon for a left-polarized photon.
If the energy distribution sums up polarizations, both $F$ and $\widetilde F$
give the same result.
Note that  since ${\cal F}$ is $CP$ even, it corresponds to Model 3 with scalar DM. 
On the contrary, the Majorana pair is $CP$ odd, and
the amplitude picks up $\widetilde{\cal F}$.

\section*{Electroweak bremsstrahlung} 
For the annihilation of Majorana DM  $\chi$ via scalar exchange, 
we derive a univeral amplitude for vector boson $V$ emission
$\chi + \chi \to f(p_1)\bar \ell(p_2)V(k,\epsilon)$
in the static limit. 
The vector boson can be $W, Z$ or $\gamma$. As $v_{\rm rel}\to 0$, 
only the composite of $\chi \chi$ with zero total angular momentum gives a contribution, 
and we get an amplitude of the form of Eq.~(\ref{amp}) with the minus sign selected, and $C$ replaced by $C_V$ (in
 which we collect the model-dependent couplings and coefficients). 
As usual, we assume that the chirality of the massless lepton field is conserved.
Note that the two terms in the numerator do not  interfere only when $k^2= 0$
(as for the photon $V=\gamma$).
QED-like gauge invariance (i.e. ${\cal M}\cdot k=0$) is satisfied
even for $k^2=m_V^2 \ne 0$. 
We calculate the annihilation cross section to be
\begin{equation}
v_{\rm rel}{d \sigma\over dx_1 dx_3}
=   {|eC_V|^2\over 16\pi^3m_\chi^2}  \    
{ [(1-x_1-\delta_V)^2+(1-x_2-\delta_V)^2-2\delta_V(1-x_3+\delta_V)](1-x_3+\delta_V)
                              \over
      (1-2x_1-r)^2  (1-2x_2-r)^2} \ ,
\end{equation}
where $r=m_X^2/m_\chi^2$ and 
$\delta_V=\fr14 m_V^2/m_\chi^2$. 
In the case of $W/Z$ emission, we assume a common mass $m_X$ for
all $SU(2)$ partners of the heavy exchanged particles.
Note that 
$x_3=E_V/m_\chi$ and $x_1$ lie in the ranges $[2\sqrt{\delta_V},1+\delta_V]$ and $[0,1-\delta_V]$, respectively.
The energy distribution of $V$ can be obtained
by integrating over $x_1$ within the limits
$x_1^\pm=(2-x_3\pm\sqrt{x_3^2-4\delta_V})/2$, and the energy distribution of $f$ can be obtained by integrating
over $x_3$ from $1-x_1+\delta_V/(1-x_1)$ to  $1+\delta_V$ .
On setting $\delta_V=0$ we recover the cross section for $V=\gamma$ and $C_\gamma$ is simply $C$
given in the photon case. The values of $C_W$ and $C_Z$ are given by
\begin{equation}
 C^2_W=C_\gamma^2 
{T(T+1)- T_3 T_3' \over 2 \sin^2\theta_W}  \ ,
\label{CV1}
\end{equation}
and
\begin{equation}
C^2_Z=C_\gamma^2
{(T_3 -Q\sin^2\theta_W)^2 \over \sin^2\theta_W \cos^2\theta_W}  \,,
\label{CV2}
\end{equation}
where $T, T_3, T_3'$ are the weak-isospin numbers of the exchanged particles and $\theta_W$ is the weak 
mixing angle. Since the spinors in Eq.~(\ref{amp}) can have either $L$ or $R$ chirality depending on the underlying physics, the weak-isospin numbers must be chosen accordingly.
The exchanged particle $S$ of Model~1 is an $SU(2)$ singlet with $Q=-1$. 

For scalar DM $\phi$ that annihilates via fermion exchange, 
a relative sign is flipped in the numerator of the amplitude, 
and 2 additional terms appear for $V=W,Z$, but not for $V=\gamma$:
\begin{equation} 
{\cal M} =- {8 eC_V\bar u_h(p_1)
( \not p_2 \not\epsilon \not k+ \not k\not \epsilon \not p_1
  +\not k \not \epsilon \not k -k^2\not\epsilon)    v_h(p_2)
\over [(p_1-p_2+k)^2-4m_X^2][(p_1-p_2-k)^2-4m_X^2] }  \,,
\end{equation}
where $h=L, R$, and $C_W$ and $C_Z$ are as in Eqs.~(\ref{CV1}) and~(\ref{CV2}).
The corresponding annihilation cross section is
\begin{equation} 
v_{\rm rel}{d \sigma\over dx_1 dx_3}
= {{|eC_V|^2\over 4\pi^3m_\phi^2}} {N   
                           \over
                               (1-2x_1-r)^2  (1-2x_2-r)^2  } \ ,
\end{equation}
$$ N=[(1-x_1-\delta_V)^2+(1-x_2-\delta_V)^2+2\delta_V(1+x_3-2\delta_V)]  
     (1-x_3+\delta_V)  $$
$$  \qquad\qquad 
+ 2\delta_V(1-x_1-\delta_V)(1-x_2-\delta_V)\,,   $$
with kinematic variables defined as before (and $m_\chi$ replaced by $m_\phi$).
For the interaction Lagrangian of Model~3 in the table,
the exchanged particle $E$ is an $SU(2)$ singlet with $Q=-1$. The realization of Model~3 in Ref.~\cite{bremm} uses 
an $SU(2)$-doublet $(N^0,E^-)$ for the exchanged particles with
$T={1\over2}$, $T_3=-{1\over2}$, $T_3'={1\over2}$ and $Q=-1$.

The difference in the $W/Z$ spectra for Models~1 and 3 arises from terms proportional to 
$\delta_V$. In both models, the energy distributions for the photon
and the $W/Z$ differ substantially due to kinematic effects if $D$ is just above the $W/Z$ thresholds, 
while for a high mass $D$ the difference is suppressed by $\delta_V$.
\\

{\it{Acknowledgments.}}
We thank T.~Weiler for a correspondence and J.~Kumar for discussions. 
DM thanks the University of Hawaii and the KEK Theory Center for their hospitality while this work was in progress.
This work was supported by DoE Grant Nos. DE-FG02-84ER40173, 
DE-FG02-95ER40896 and DE-FG02-04ER41308, by NSF
Grant No. PHY-0544278, and by the Wisconsin Alumni Research Foundation. 


\end{document}